\documentclass[preprint]{aastex}

\slugcomment{Submitted to Astrophysical Journal}

\def\etal{{\it et al.} }                
\def\kms{km~s$^{-1}$}
\def\GTT{Tenorio-Tagle}
\def\Cas{Mu\~noz-Tu\~n\'on}
\def\h2{H\,{\sc ii}}
\def\ha{H$\alpha$ }
\def\2ps{[S\,{\sc ii}]}
\def\3po{[O\,{\sc iii}]}
\def\o3ha{[O\,{\sc iii}]/H$\alpha$ }
\def\s2ha{[S\,{\sc ii}]/H$\alpha$ }

\begin{document}

\title{ON THE ONGOING MULTIPLE BLOWOUT IN NGC 604}

\author{Guillermo Tenorio-Tagle}
    \affil{INAOE, Apartado Postal 51, Puebla, Pue. MEXICO}
\author{Casiana Mu\~noz-Tu\~n\'on}
    \affil{Instituto de Astrof\'{\i}sica de Canarias, 38200 La Laguna, Tenerife, SPAIN}
\author{Enrique P\'erez}
    \affil{Instituto de Astrof\'{\i}sica de Andaluc\'{\i}a (CSIC), Aptdo. 3004, 18080 Granada, SPAIN}
\author{Jes\'us Ma\'{\i}z-Apell\'aniz}
\affil{Space Telescope Science Institute, 3700 San Martin Drive, 21218 Baltimore, USA}
\affil{Laboratorio de Astrof\'{\i}sica Espacial y F\'{\i}sica Fundamental-INTA,
Apartado Postal 50727, 28080 Madrid, SPAIN}    
\author{Gustavo Medina-Tanco}
    \affil{Instituto Astron\^omico e Geof\'{\i}sico, University of S\~ao Paulo, BRAZIL}

\begin{abstract}
Several facts regarding the structure of NGC 604 are examined here. 
The three main cavities, produced by the mechanical energy from massive
stars which in NGC 604 are spread over a volume of 10$^6$ pc$^3$, 
are shown here to be undergoing blowout into the halo of M33. 
High resolution long slit spectroscopy is used to track the impact from massive stars  
while HST archive data is used to display the asymmetry 
of the nebula.

NGC 604 is found to be a collection of photoionized 
filaments and sections of shells in direct contact with the thermalized 
matter ejected by massive stars.  
The multiple blowout events presently  drain the energy injected by massive stars and thus the densest photoionized gas is found almost at rest and is expected to suffer a slow evolution. 

\end{abstract}

\subjectheadings{{\bf ISM}: kinematics and dynamics -- H\,{\sc ii} regions -- NGC 604}

\section{INTRODUCTION}

Upon the formation of massive stellar clusters, the energy input 
from massive stars leads to a rapid structuring of the surrounding 
medium. This was originally thought to be due to the high pressure 
acquired by the constant density photoionized gas which, upon the 
formation of the H\,{\sc ii} region, would then expand with a speed
smaller than the sound 
speed ($c_{HII}$ $\sim10$ \kms) onto the surrounding neutral gas to acquire a 
radius

$R_{HII}(t) = R_{St} (1 + 7/4 (c_{HII}t/R_{St}))^{4/7}$

\noindent where $R_{St}$ is the Str\"omgren radius 
(Spitzer 1968; Osterbrock 
1989). Much larger speeds were envisaged in ''champagne flows'' as 
a result of the pressure gradient between the photoionized parent 
cloud and the photoionized inter-cloud medium, suddenly established 
as the ionization front (either D or R type) overruns the parent 
cloud edge (\GTT\ 1979). In this case the ionization front rushes 
to infinity and strong isothermal shocks develop within the photoionized 
region.  The shocks are 
driven by the ionized cloud matter that supersonically 
streams into the intercloud 
medium in an attempt to lift and equalize the pressure between the 
two regions. The champagne phase would then last until the pressure 
gradient is smoothed out; fact that is accomplished only once the 
whole ionized cloud is dispersed.

A completely new paradigm resulted from taking into consideration 
the mechanical energy produced by massive stars and the evolution 
of superbubbles. In this case three concentric distinct regions were 
shown to develop around the exciting source (Castor et al. 1977): 
The central region ({\it a}) contains the hypersonic wind from the 
massive star(s). At a certain radius, $R_a$, this is abruptely 
thermalized at a reverse shock. Given the speed of the wind, the high 
temperature ($T$ = 1.4 $\times$ 10$^7$ ($v_{wind}$/1000 
\kms)$^2$ K $\sim$ 10$^6$--10$^7$ K) and high pressure acquired upon 
thermalization, allow the shocked wind to fill a large volume (region 
{\it b}), while driving a strong shock into the surrounding gas. 
The latter shock then sweeps and accelerates the surrounding ISM into 
a dense expanding shell (region {\it c}), that grows as a function of 
time as (Castor et al. 1977; Weaver et al. 1977)

$ R_s(t) = 0.76 (\dot{E}/\rho)^{1/5} t^{3/5} $

\noindent where $\dot{E}$ is the mechanical energy input rate and $\rho$ 
the mass density of the background medium.
Castor \etal (1977) also showed that, given the number of recombinations 
within the expanding shell and its rapid growth, the ionization front 
became trapped within the shell early in the evolution of the 
system. This was shown to happen despite the fact that most of the 
volume is filled with the hot thermalized wind, which is 
a transparent medium to the ionizing photon flux due to its high temperature. 
Consequently, the H\,{\sc ii} region gas in this model was shown to be 
rapidly confined to a narrow 2D structure within the expanding shell.

Other models, accounting only for the mechanical energy deposited by OB
associations and starbursts (\GTT\ \& Bodenheimer 1988), have clearly 
shown how the shells of swept up matter can easily be disrupted upon 
crossing a large density gradient. This phenomenon, commonly referred 
to as blowout, occurs as a shell suffers a sudden acceleration, fact 
that drives it Rayleigh-Taylor unstable and causes its rapid fragmentation.
The event then leads to the venting of the hot superbubble interior 
onto the low density gas, where a new shell of halo swept up matter will
develop. More recent studies have also shown that blowout also leads 
to the escape of the $uv$ photons, which otherwise would be trapped 
within the decelerating shell. The escape of ionizing radiation has 
been shown to lead to the development of an extended, low surface 
brightness conical H\,{\sc ii} region ahead of the superbubble leading 
shock (\GTT\ et al. 1999).

Here we center our attention on the nearest northern hemisphere 
giant \h2 region NGC 604. The giant nebula is the largest in the 
neighbouring spiral galaxy M33 (at a distance of 840
 kpc; Freedman \etal 1991), and is located in one 
of the spiral arms,  2.4 kpc from the galaxy centre. 
Looking at NGC 604, it is quite evident that it is not a spherical
nebula.  It presents a large number of shells, filaments or arches
of a wide variety of lengths and intensities.  Its general appearance 
is, however, very similar to other giant H\,{\sc ii} regions, in particular
to the 30 Doradus nebula. The large number of shells and networks of 
shells in 30 Doradus have been identified spectroscopically by Chu \& 
Kennicutt (1994). Such a structure, i.e. expanding shells within shells, 
imply a very inhomogeneous rapidly structured, or pre-structured, 
parent cloud density distribution. An endless number of tunnels, 
or cavities, across which the thermalized winds from the massive stars 
could freely flow causing their expansion and further growth, as to in
projection allow for the appearance of a network of nested shells.
In such a case, the thermalized ejecta, the hot coronal gas, is to
be in direct contact with the H\,{\sc ii} region gas, and the latter 
would be confined to the narrow denser walls of the network of tunnels 
or to the sections of the expanding shells, presenting thus a 
three dimensional superposition of multiple 2D structures.

Here we analize the ionization structure, and the velocity field of
NGC 604, with the aim of 
unveiling its true morphology and possible evolution of the nebula. 
In section 2 we analyse the two data sets used in this work, and
in section 3 we present a discussion of the results and our conclusions.

\section{Analysis and results}

\subsection{$HST$ imaging}

We have retrieved from the $HST$ archives narrow band images in the 
emission line filters of \ha\ (F656N), [O\,{\sc iii}] $\lambda$5007 (F502N), 
and [S\,{\sc ii}] $\lambda$6717+6731 (F673N). 
Each final image is the combination
of two 1100 s exposures, using the {\sc STSDAS} tasks {\tt crrej} and 
{\tt wcombine}. Figure 1 shows the image ratios 
corresponding to \o3ha\ and \s2ha.

A cursory look at these two image ratios shows that, as expected, 
they trace distinct different regions of the nebula. The \s2ha ratio 
traces the low excitation regions and is not affected by extinction.
The \o3ha ratio maps the high excitation regions, but it can be affected
by extinction, so that a high value always indicates high excitation,
but a low value can be a combination of both low excitation and/or
high extinction. This technique of using narrow band emission line 
imaging ratios has been used successfully in the literature to study
the ionization structure and radiation field in different types of 
photoionized sources, such as Seyfert and Starburst galaxies, where ionization
cones and other collimated structures have been thus uncovered 
(Tadhunter \& Tsvetanov 1989; P\'erez et al. 1989). 

The high excitation O$^{++}$ region is maintained by radiation with at 
least 35 eV, while the low excitation S$^{+}$ zone requires only of 
photons more energetic than 10 eV. Thus, in an idealized constant density
\h2 region, with a source of $uv$ photons arising from
the center of the nebula, the S$^{+}$ zone should be
concentric and surrounding the O$^{++}$ zone. This is clearly not the 
case in NGC 604 (see Figure 1). 
Instead, although the high excitation region is 
largely confined by the outer low excitation region, there exists a 
large, almost straight dividing line that spans across $\sim$ 200 pc 
(50 arcsec) almost through the center of the nebula (offset by some 
10 arsec), running from the extreme north to the southern edge. 
The sharp boundary between the two emitting volumes (hereafter referred 
to as the ''ridge''), has no explanation within the standard case. Note
however, that (unlike R136 in 30 Doradus) the ionizing cluster in NGC 604 
is not centrally concentrated, but rather the 186 massive O stars 
visible in the $HST$ images are here distributed over a projected area 
of about 10000 pc$^2$ (Hunter et al 1995; Terlevich et al 1996; 
see figure 2 in Gonz\'alez Delgado \& P\'erez 1999). 
Furthermore, from a recent radio continuum 
and {\ha} study, Churchwell \& Goss (1999) conclude that the main
emission centres (their knots A-F) are also photoionized from the 
inside by heavily obscured massive stars, and an average of five or 
more O stars are required to produce the ionization of each of the 
condensations. These come to enhance the number of massive O stars 
to more than 220, scattered over the projected central 10000 pc$^2$ area. 
Also from this work an extinction map was obtained with values ranging 
from 0.3 to 0.5 magnitudes, thus implying that we see most of the 
nebula without obscuration (see also D\'\i az et al. 1987). 
Another recent analysis (Ma\'{\i}z-Apell\'aniz et al. 2000) finds a discrepancy
between the extinction measured from the Balmer decrement and the one
found by Chuchwell \& Goss (1999), suggesting that a large fraction of the
obscuring dust is mixed with or located close to the gas. This together with 
the CO detection (Viallefond \& Goss 1986) implies the existense of a molecular 
cloud behind the photoionized nebula NGC 604.  

The length of the ridge between the two different excitation regions demands 
the existence of a sharp discontinuity across which distinct physical 
conditions operate and is most likely caused by the existence of a molecular
wall (Wilson \& Scoville 1992, Ma\'{\i}z-Apell\'aniz et al. 2000). The ridge, in
deep \ha images seems to result from the chance
alignment of the three main shells, or superbubbles, whose sharp
and elongated edges coincide to give the appearance of a long ridge
(see Figures  1 and 2).

Further information regarding the structure of the region can be
derived from the kinematics of the ionized gas.  In what follows we
analyse spectra taken at the WHT (4.2m) at the Observatorio del
Roque de los Muchachos using the double arm ISIS spectrograph.

\subsection{The peak intensity features}

A 1 arcsec wide slit was placed east-west at ten different locations 
stepped by 2 or 3 arcsec across the nebula (the step between slit positions
\#6 and \#7 is 2 arcsec).  
The sampling along the spectral and spatial direction in the slit is
0.37 \AA\ $\rm pixel^{-1}$ and 0.33 arcsec $\rm pixel^{-1}$, 
respectively.
The precise location of the slits is shown in Figure 2 and a
detailed description of the observational set-up and data
reduction can be found in Mas-Hesse et al (1994), Ma\'{\i}z-Apell\'aniz
(1999), and Ma\'{\i}z-Apell\'aniz et al (1999). Figures 3a to 3j
display the measured velocity and flux of the $H{\alpha}$ emision
line along each of the slits crossing the nebula. 
The measurements have been made by fitting a single Gaussian
component to the observed {\ha} emission line, except where a good fit
to the line required two or sometimes three components.
The velocity field (\kms) is derived from the center of the Gaussian fits,
referred to the {\ha} rest wavelength ($\lambda_0=6562.58$ \AA).
The line intensity scale (in arbitrary units) 
gives an indication of the density, or the column density at the
observed locations. The median velocity is $256\pm9$ \kms.

The long ridge that divides the east and west sides of the nebula
is evident in all the slits.  The feature presents a maximum
intensity at pixel $\sim300$ on all slits.  The central
velocity at such local maxima is very uniform with values of -250
\kms to -260 \kms across the ten slits.  These are very similar to
the galaxy local reference frame velocity.  The sharp ridge across
the nebula is most probably a high density wall with a very steep
drop in intensity towards the east, noticeable on all slits.  Much
further away (around location 370) one finds the Big Loop.  This is
easily identified in Figure 2 across slits 1 to 5 (see insert sketch).

Further out on slit 1, at positions 430 and 445, the two small
local maxima correspond to two small \h2 regions (Figure 2) and
both also present a velocity of -260 \kms.  The two point-like
\h2 regions have been used as reference for the positioning of the
slits.

Churchwell \& Goss (1999) have identified six radio continuum maxima
and labelled them as A--F.  In what follows we shall keep their
nomenclature (see Figure 2).  The brightest zones of the nebula are
crossed by slits 6 to 10.  The sharpest peak is knot A easily found
on slit 8 at pixel 300.  Its spatial extent, measured as the apparent
width at half intensity, is about 5 pixels, or 1.8 arcsec,
equivalent to 7.9 pc, if one assumes a distance scale of
4.1 pc arcsec$^{-1}$.  Knots A and B (on slits 8, 9 and 10) lie on 
top of a high density wall and in fact between them conform the most
evident southern part of the long ridge.

The other main emitting knots are also clear in our spectra.  Knot
D is very elongated and most evident on slits 4 to 7.  It is
brighter on slit 6 (with 10$^4$ counts at position 245)
and of comparable brightness on slit 7 (9000 counts at 250). 
The intensity of this features slowly fades through slits 5
(5500 counts at 245), 4 (3200 counts at 240), 3 (2300
counts at 245) and 2 (1700 counts at 240), and it is absent on
slit 1.  A deep inspection of the image shows that in fact, the
feature ends up running almost parallel to slit 2 (from position 
250 to 280).  Therefore, although along the slit direction, knot D is
relatively narrow ($\sim15$ pixels wide at half intensity on
slit 7), it extends continously from the north-west northwards and
then to the east following a concave curvature.  The emission
increases once again as one moves towards the south across slits 3--7 
(positions 280--290).  Towards the south the feature
acquires large intensity values, similar to those found along the
extension of knot D towards the north.  If one follows this
roundish shape moving onto slit 8 one finds the next maximum in
intensity (knot C).  In our spectra this splits into two distinct
knots, each of them with typical sizes of 10 pixels ($\sim3.7$ arcsec
width at half intensity), equivalent to 15.2 pc.

With the exception of knots A and B, the intensity maxima 
described above clearly define a central zone with a much lower
\ha intensity.  There, the flux measured in between maxima
reaches the lowest possible values (as on slits 4 and 5).  There
seems to be an empty hole, the boundary of which is well delineated
by some of the main emitting knots and by the slightly curved
central section of the ridge on the east side.  We have then
identified an elliptical shape cavity with a diameter $\sim40$ pixels 
($\rm\sim15~arcsec \times 11.25~arcsec$ or 61.5 pc $\times$ 46.2 pc).

Within the central hole we identify a small but extended zone
(the ''intrusion''), which extends (E--W) from pixel 300 to 265 along slit 6, 
i.e. some 15 pixels (22.75 pc) while reaching a maximum flux of 2500 units.

\subsection{The velocity field}

The two reference \h2 regions (at pixels 435 and 445 on slit 1), both ends of 
every slit and all obvious intensity maxima traversed by the slits (with the 
exception of the intrusion in the main cavity) show the same velocity,
$-255\pm 10$ \kms. This value, has then been assumed as the
galaxy reference frame.

The velocity curves also display a different trend between the \2ps 
and the \3po dominated regions.  There is no zone within the east
side of the \2ps region that shows line splitting.  All slit
positions within this \2ps dominated region, regardless of the
\ha intensity, are well fitted with a single Gaussian
with peak velocities about -260 to -280 \kms, all
within 10 \kms\ from the rest-frame velocity of M33.  \\
 
One can therefore conclude that there are no signs of this region
being affected by the mechanical energy from the massive stars.
The \2ps zone presents in fact evidence of a slow radial expansion
with respect to the M33 reference frame velocity.  Note also that
not even the largest loop of NGC 604, apparent at all wavelengths
at the outer edge of the nebula (at the extreme North-East edge,
across slits 1--6), presents any sign of line splitting and thus
we regard it as an old structure at the edge of a molecular
cloud.  The exact opposite is true within the \3po dominated
zone.  There the velocity field, through large line splitting,
evidences the mechanical power from massive stars, capable of
displacing the ISM as it generates giant cavities.

Apart from the obvious main cavity, there are also several smaller
less obvious holes, the borders of which are also traced in the
$HST$ \ha image.  The largest ones (cavities 2 and 3, see
Figure 2) are to the south and north, immediately adjacent to the
rim of the central hole, and are bound on their eastern side by the
ridge.  The two of them fade away in opposite directions, cavity 2
appears open and fainter towards the south (see slits 8--10 at pixels
260 to 300), while cavity 3 towards the north (see slits 1--3 at pixels
280 to 300).  Other even smaller and fainter features are even less
apparent and it is a difficult task to describe them with the
same accuracy.

Line splitting has been detected in all three cavities.  The
largest inferred expansions from the spectroscopic data are $\sim$
100 \kms\ with respect to the galaxy rest frame.  In the regions of
line splitting, the two line components are displaced from the
galaxy rest frame velocity.  The redshifted component acquires in
all cases a maximum value $\sim$ -240 \kms.  It is also the
redshifted component, with exception of that arising from the
intrusion feature in the main cavity, the one that invariably
presents the largest intensity.  On the other hand, 
the blue shifted components reach values of -370 \kms, implying expansion
speeds of up to 100 \kms\ with respect to the galaxy rest frame.
The blue shifted components present also in all cases a low
intensity flux, always below 100 counts (see Figure 3), except in
the intrusion feature seen in projection within the main cavity.
The ''intrusion'' thus appears as a dense feature (intensity $\sim$
2000 counts at pixel 270, on slit 6), blueshifted some 20 \kms\ with
respect to the galaxy rest frame velocity.  At the same position
the redshifted counterpart reaches speeds $\sim$ -220 \kms, very
similar to those detected at the same pixel location on the
neighbouring slits 5 and 7.  There are also some regions, very
few, which display more than two components.  These are all found
within the main cavity.  A precise modeling of these is difficult
to do, particularly when the measured parameters are affected by
large uncertainties.  All one can assure is that more than two
Gaussian components are required to reproduce the emission line profile.  We
must centainly be intercepting remnant shells or low intensity
filaments along the line of sight, stressing the idea of depth
or of a 3D structure of the nebula.

Note however that the velocity centroids of the splitted lines do
not conform, in any of the studied cases, to complete velocity
ellipsoids, as those expected from an ionized shell
expanding into a uniform density medium;  i.e., an ellipsoid showing
maximum speed of approach and recession when looking across the
centre of the cavities, steadily merging with the galaxy
rest-frame velocity as one approaches the borders of the expanding
structure (see, e.g. Ma\'{\i}z-Apell\'aniz et al. 1999).  The data  instead 
shows distinctly open structures.  Line splitting is clearly
found even when looking over some of the borders, or rims, of the
most obvious cavities (see slits 2 and 3).

Figure 4 maps the velocity centroids of the two main line
splitting components across the north--south direction and
through the centres of cavities 1, 2 and 3.  Figure 4a shows the
range of velocities detected at pixel 270.  
Figures 4b and c, show the trend of the blue and the red shifted
components in cavities 2 (pixel 285) and 3 (pixel 290), 
respectively.  In the three cases the velocity
ellipsoids expected from shell-like structures expanding onto the
parent cloud are not recovered.  The line spliting shown in
all of them leads to widely separated velocity features with a
$\Delta v$ in the range of 60--70 \kms.  Following the slits
across the cavities, i.e., along the east--west direction (see
Figure 3), one can see that the line splitting is confined by the
borders of cavities.  Although slit 2 at position $\sim$ 280 
presents evidence of a large line splitting right at the 
high intensity border between cavities 1 and 3. This fact  
indicates that  
the blue shifted outflows from these cavities have already managed to 
surpase the dimensions of the cavities that originally confined them.  The
trend detected in the three main cavities of NGC 604 thus
resembles the trace left by the cones that channel superwinds
out of starbursts galaxies, as in NGC 253, and M82 (see Heckman
et al. 1990).  NGC 604 being a much smaller and weaker source,
represents a small-scale version of a superwind event.  However,
it is a clear example of the venting of the hot superbubble
content into the halo of M33, as well as of the streaming or leakage 
of $uv$ radiation into the halo of M33.

Positions with line splitting associated to cavities 2 and 3 are
also found at locations far from  the main extended region of
star formation.  The line splitting at these positions appears to
coincide with radial gaseous filamentary regions at the edge of
the emitting zone.  The structures thus resemble the tracing of
conical outlets through which the hot gas, which otherwise could
be used to power the expansion of bubble-like structures, freely
expands away from NGC 604, most probably into the low density halo
of M33.

Typical of the splitted lines, is their low intensity fast blue
shifted expanding components, much faster than those detected in
the redshifted counterparts.  Also, note that the blue shifted
components present in all cases, but in the intrusion, a much
lower intensity, indicative of a recent sudden burst into a lower
density medium.  Both of these arguments together with the fact
that the blue shifted velocity component is also seen on top of
the projected boundaries of the cavities (as in slits 2 and 3),
support the evidence of an ongoing blowout.  In this scenario, the
intrusion feature corresponds to the densest remaining part of a
shell that has undergone blowout in our general direction.  It
seems also relevant to point at the elongated rim of cavity 3
which may have developed during blowout, and that spans across to
conform the northern half length of the long ridge.

\section{Discussion and Conclusions}

Central to the study of \h2 regions is to define the way they
evolve and for this an accurate description of their morphology is
a key issue to confront with theoretical models.  
NGC 604 is excited by a widely spread
cluster of some 200 massive stars, and these  have produced two
non-concentric different excitation zones distinctly different
also from a kinematical point of view.  Our kinematical data
provides also strong evidence of an ongoing multiple blowout of
the photoionized nebula into the halo of M33.

The \3po dominated section of NGC 604 presents a corrugated
multiple shell appearance, presently bursting along several
directions.  This is most evident across bubbles 1, 2, and 3 and
along the filamentary channels detected at large radii, both along
the north and southern edges of the nebula.  Note that bubbles 2
and 3 spread out far from the extended cluster of massive stars.
Furthermore, in all these cases line splitting consistently shows
the densest (high intensity) gas only slightly redshifted with
respect to the galaxy reference frame, and thus it is expected to
evolve very slowly during the life time of the exciting stars.  On
the other hand, the low intensity blueshifted counterpart is
clearly affected by a rapid evolution, as expected from the
surface of superbubbles bursting towards us at high speeds into
the halo of M33.

The eastern \2ps section of NGC 604 presents a completely different
structure that seems kinematically unaffected by the winds and
supernovae from the extended exciting cluster.  This conclusion
arises from the absence of line splitting and the lack of
bubble-like structures.  The \2ps region in fact gives the
impresion of a large photoionized section of the surface of an
extended cloud expanding away from us with a velocity smaller than 10 \kms.

Many models have been used to explain diffuse ionized nebulae (see
section 1).  However, to account for the observed facts above, a
particular scenario is required.  This involves a pillar
morphology of the parent molecular cloud, where massive star
formation must have been triggered at one end and deep along its
surface, perhaps by a major disturbance such as the M33 spiral
density wave.  The star formation event covers in projection an
area of some 10500 pc$^2$ and must probably occupy a volume
$\sim$ 10$^6$ pc$^3$.  We envisage star formation as to have
occured in many places deep enough inside the cloud, and as
closely spaced together, as to allow the mechanical energy from
massive stars to build up a network of twisted tunnels, filled
with the thermalized ejecta; in a similar manner to that envisaged
by Cox \& Smith (1974) for OB associations in the galaxy.  The
tunnels will then traverse the tip of the giant molecular cloud
and become broader, particularly in the outward direction; i.e. 
away from the symmetry axis or densest part of the cloud, in the
places where more energy is deposited per unit time, or where the
local surrounding cloud density is the lowest.  In
some of these places the energy from the ''super-tunnels'' will even
burst out of the cloud, releasing the hot coronal gas into the low
density surrounding medium.

Looking at the cloud along its symmetry axis, the enhanced
diameter sections of the cylindrical tunnels would appear like the
borders, or rims, of individual ``shells'', and the collection of
protuberances along the surface of a tunnel could cause the
apearance of networks of nested shells.

NGC 604 has three large cavities evident in the $HST$ \ha
image.  The rims of these conform the broad borders of cavity 1
and of the neighbouring more distorted ``bubbles'' 2 and 3.  The
alignment of the three borders, jointly compose the dense ridge
that separates the high from the low excitation regions of the
nebula.  Note that the end of cavity 3 is the most distorted of
them all, and extends for some distance away from the center of
star formation.  The walls of the three cavities must extend deep
into the cloud to inhibit in this way an important leakage of hard
$uv$ radiation into the \2ps main emission section of the nebula.

From the kinematic data we know that the three main bubble
structures have burst into the halo of M33, as evidenced by the
line splitting across their projected cross-section.  This confirms the 
detection of the main bubbles through
a completely different observational approach using Fabry-Perot
data (see Mu\~noz-Tu\~n\'on et al. 1996).  The cavities, or tunnels, may have
also burst in other directions, leading to the giant faint loop
structures, and filamentary structures seen at large projected
distances from the star forming zone.  These regions being far
from the collection of exciting stars present thus a low
ionization parameter and are most evident in \s2ha.

Clearly a detailed model of  NGC 604 should strongly depart from the
idealized one dimensional spherical symmetry, while allowing
photoionization and the stellar wind energy input to jointly work
on the shaping of the parent cloud.  The above thus implies that:

a) In NGC 604 the formation of massive stars has taken place
within a large volume ($\sim$ 10$^6$ pc$^3$), and the mechanical
energy from these has rapidly structured the parent cloud into a
collection of interconected bubbles and tunnels filled with the
thermalized matter ejected in winds and eventually SN explosions.

b) Consequently, the \h2 region is not filled with photoionized
gas and thus it cannot possibly expand as originally thought \h2
regions would do (see section 1).  The matter surrounding massive
stars has rapidly been confined to the borders or walls of the
multiple bubbles and tunnels, and there, exposed to the $uv$
radiation causes the multiple shell or filamentary appearance
characteristic of giant \h2 regions.

c) The hot coronal gas that initially drives the expansion of the
network of bubbles and tunnels moves rapidly across the organized
network to remain in direct contact, and thus at the same
pressure, as the high excitation photoionized section of the
nebula.

d) The evidence of blowout into the halo of M33, also implies that
the expansion velocity of tunnels and shells should rapidly
plummet as the driving energy escapes into the halo of the galaxy.
In this case, the general structure is expected to suffer little
changes during the remaining life-time of the exciting stars.

e) Blowout should allow the hot coronal medium to engulf the whole
cloud and to establish a new pressure balance along the extended
cloud surface.

f) Far away photoionized sections of the cloud would inevitably
suffer their disruption through champagne flows, although details
of such flows once the cloud surface is also permeated by the hot
coronal medium, remain to be worked out.

g) As a consequence of blowout, an important and increasingly 
larger amount of $uv$ photons escape the nebula. This should
largely contribute to the ionization of the extended halo of M33.

{\bf Acknowledgements}
We gratefully acknowledge the hospitality of the Guillermo Haro
program for advanced astrophysical research (workshop 1999) 
which largely
promoted our collaboration in this project.
We are also grateful to the CONACYT (M\'exico) -
DGES (Spain) proyect 97-0158, and CSIC-CONACYT project 99MX0021
as well as the CONACYT project 28501-E (M\'exico), and FAPESP/CNPq (Brazil). 
We acknowledge the hospitality of the Instituto de Astrof\'\i sica 
de Andaluc\'\i a (CSIC, Granada, Spain) and the Institute of 
Astronomy (Cambridge, UK).
Our thanks also to David Carter for his help with gathering of the 
long-slit data within the GEFE collaboration and to Miguel Mas-Hesse 
for his making sure we could count on very well calibrated data.

\references

\noindent  Castor J., McCray R., \& Weaver R., 1975, ApJ, 200, L107

\noindent Chu, Y.-H., \& Kennicutt, R.C.Jr., 1994, ApJ, 425, 720

\noindent Churchwell, E. \& Goss, W. M., 1999, ApJ 514, 188

\noindent Cox, d. p. \& Smith B. W., 1974, ApJ 189, 105

\noindent D\'{\i}az, \'A.I., Terlevich, E., Pagel, B.E.J., V\'{\i}lchez, J.M.,
Edmunds, M.G., 1987, MNRAS, 226, 19

\noindent Freedman, W.L., Wilson, C.D. \& Madore, B.F., 1991, ApJ, 372, 455

\noindent Gonz\'alez-Delgado, R.M., et al., 1994, ApJ, 437, 239

\noindent Gonz\'alez-Delgado, R.M., P\'erez, E., 1999, MNRAS, submitted

\noindent Heckman, T.M., Armus, L. \& Miley, G.K., 1990, ApJS, 74, 833.

\noindent Hunter, D.A., Baum, W.A., O'Neil, E.J.Jr., Lynds, R., 1996, ApJ 456, 174

\noindent Ma\'{\i}z-Apell\'aniz, J., 1999, Ph.D. Thesis, Univ. Complutense, Madrid.

\noindent Ma\'{\i}z-Apell\'aniz, J., Mu\~noz-Tu\~n\'on, C., Tenorio-Tagle, G. \&
Mas-Hesse, M., 1999, A\&A, 343, 64.

\noindent Ma\'{\i}z-Apell\'aniz, J., P\'erez, E. \& Mas-Hesse, M., 2000, in
preparation.

\noindent Mas-Hesse, M.,  Mu\~noz-Tu\~n\'on, C., V\'{\i}lchez, J. M., 
Casta\~neda, H. O. \& Carter, D., 1994, in {\it Violent Star Formation: From 
30 Doradus to QSOs}, ed. Tenorio-Tagle, CUP, p. 125.


\noindent \Cas, C., \GTT, G.,  Casta\~neda, H. O. \& Terlevich, R. 1996, AJ, 112, 1636

\noindent Osterbrock, D. E., 1989, in Astrophysics of Gaseous Nebulae and
Active Galactic Nuclei, University Science Books, Mill Valley.

\noindent P\'erez, E., Gonz\'alez-Delgado, R.M., Tadhunter, C.N., Tsvetanov, Z., 
1989, MNRAS 241, 31P

\noindent Spitzer L.: 1968: in Diffuse Matter in Space, Interscience 
Publishers. John Wiley \& Sons Inc. 
New York, London, Sydney, Toronto.

\noindent Tadhunter, C.N., \& Tsvetanov, Z., 1989, Nature, 341, 422

\noindent Tenorio-Tagle, G., 1979, A\&A, 71, 59

\noindent Tenorio-Tagle, G. \& Bodenheimer, P., 1988, ARA\&A, 26, 145

\noindent Tenorio-Tagle, G., Silich, S., Kunth, D., Terlevich, E., Terlevich, R., 1999,
MNRAS, 309, 332

\noindent Terlevich, E., D\'\i az, \'A.I., Terlevich, R.J., Gonz\'alez-Delgado, 
R.M., P\'erez, E., Garc\'{\i}a Vargas, M.L., 1996, MNRAS 279, 1219

\noindent Viallefond, F., \& Goss, W.M., 1986, A\&A, 154, 357

\noindent Weaver R., McCray, R., Castor, J., Shapiro, P. \& Moore, R., 1977
ApJ, 218, 377

\noindent Wilson, C. D. \& Scoville, N., 1992, ApJ, 385, 512

\endreferences

\vfil\eject

\begin{figure}
\centerline{\includegraphics*[angle=270,width=\linewidth]{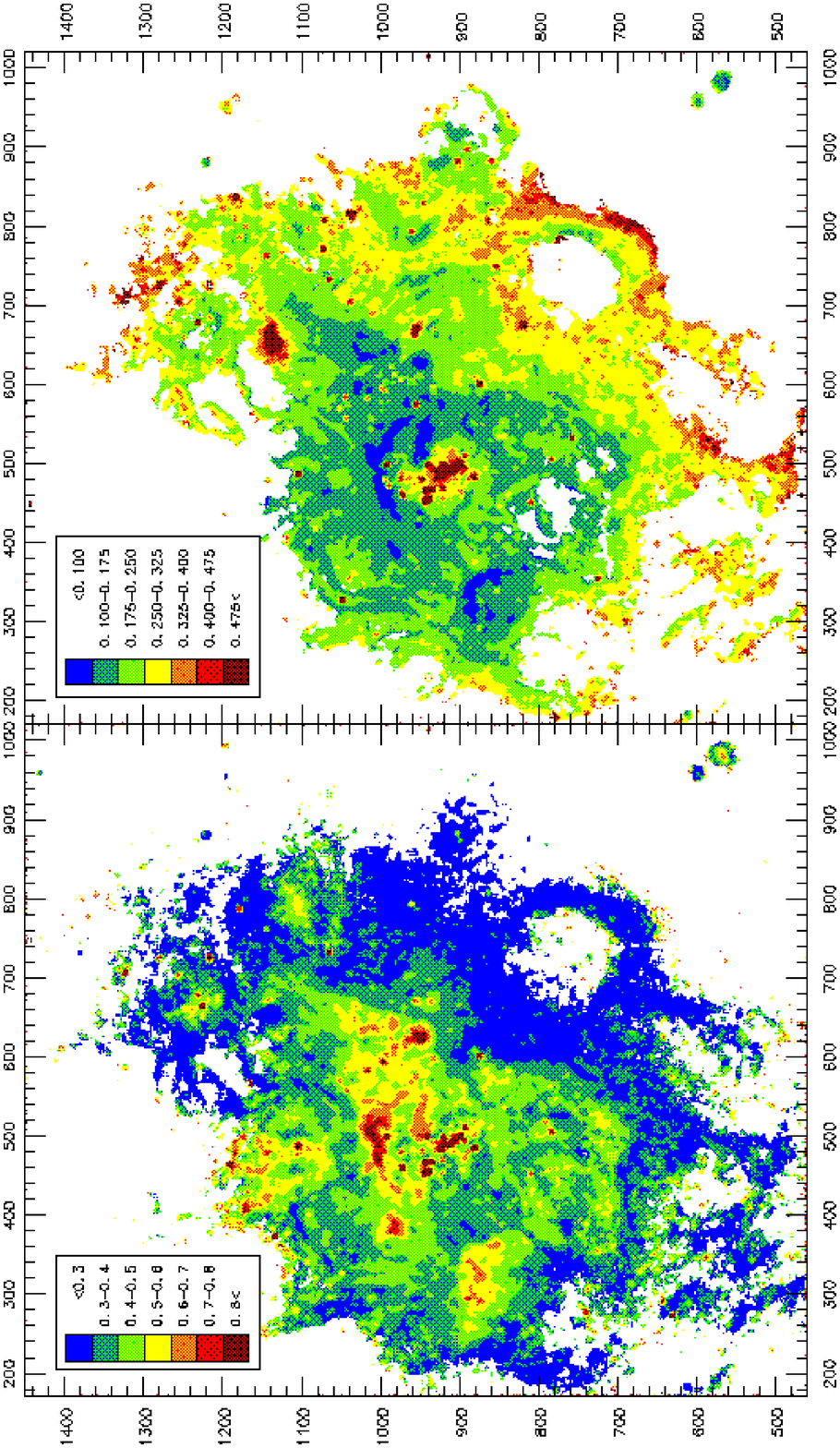}}
\figcaption{\o3ha (left) and \s2ha (right) excitation maps derived from the
$HST$ WFPC2 images of NGC 604. The scale used for each of the ratios is 
indicated in the legend. Note that for \o3ha high values correspond to high 
excitation while for \s2ha high values correspond to low excitation. The axes 
are labelled in WF pixels, with 1 pixel $\approx$ 0\farcs 1. North points
towards the lower left corner, forming an angle of 27\fdg 72 with the
vertical.}
\end{figure}

\begin{figure}
\centerline{\includegraphics*[angle=270,width=\linewidth]{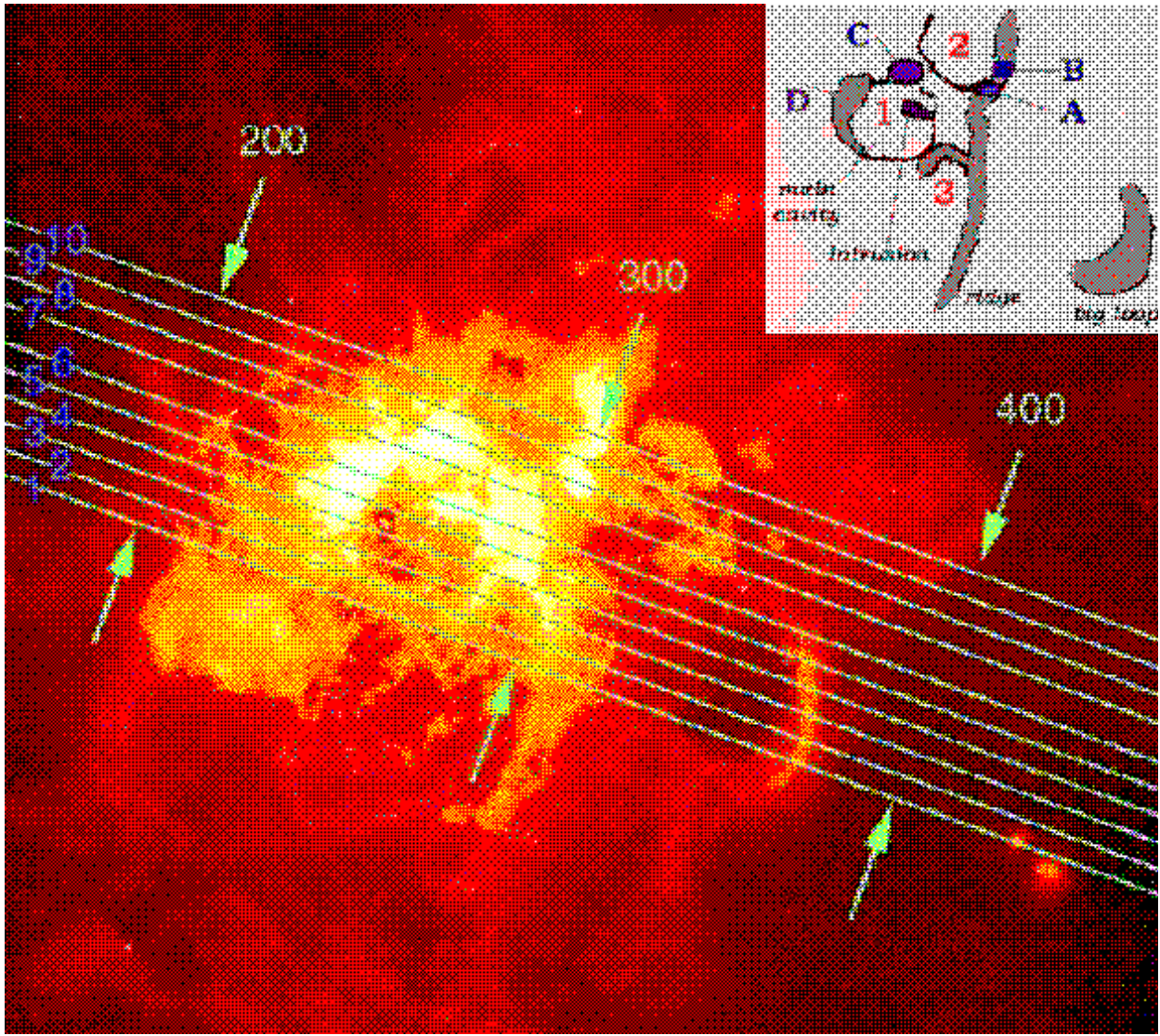}}
\figcaption{NGC 604 $HST$ \ha image with overlaid slit locations. The inset 
sketches the main features discussed in the text. }
\end{figure}

\begin{figure}
\centerline{\includegraphics*[width=.85\linewidth]{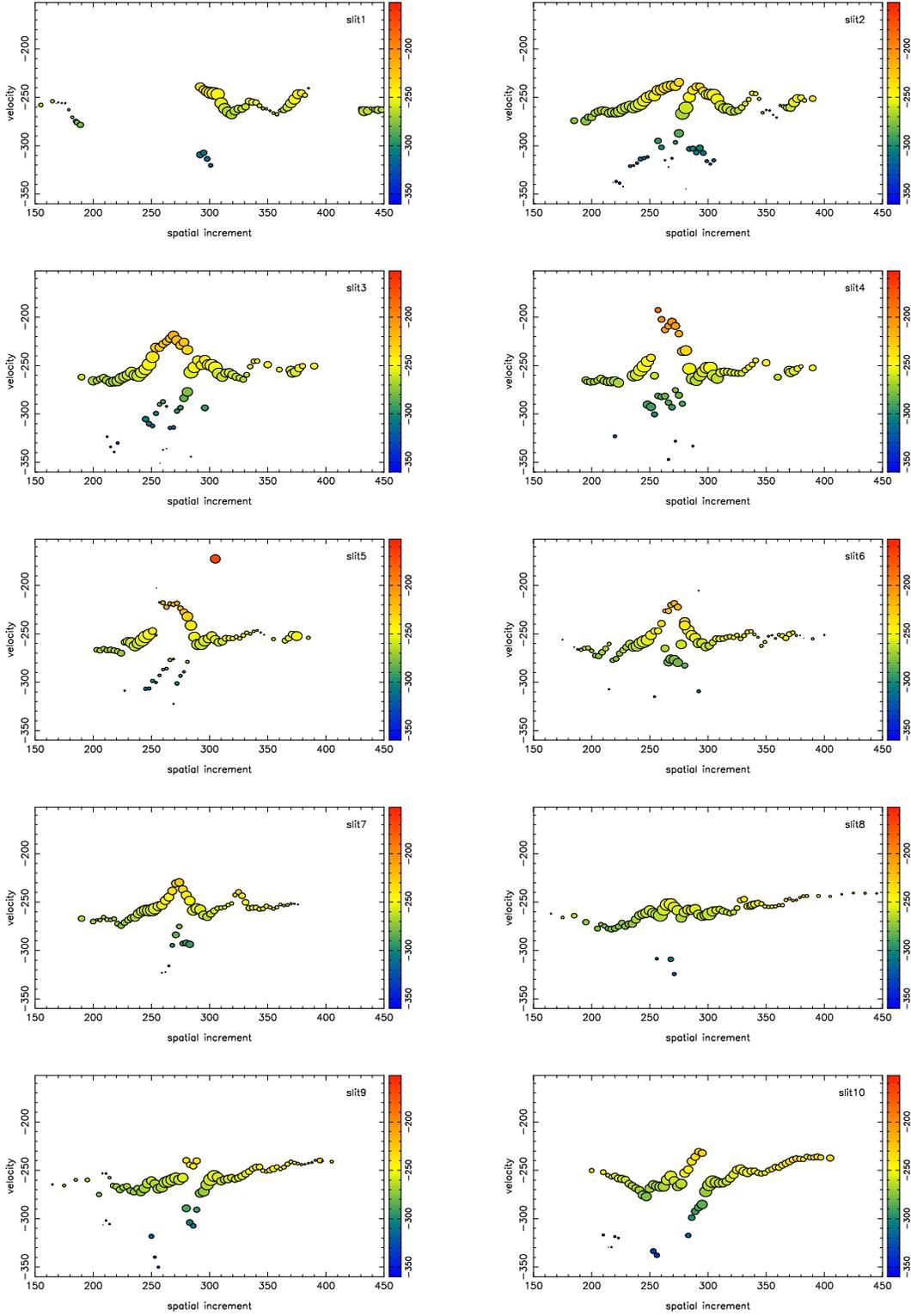}}
\figcaption{\ha velocity along the ten slit positions across the nebula. The
size of the symbol is proportional to the log of the \ha\ flux within that slit 
position. The vertical scale is in km s$^{-1}$ and the horizontal scale is in 
long-slit pixels (1 pixel $\approx$ 1.3 pc).}
\end{figure}

\begin{figure}
\includegraphics*[width=.90\linewidth]{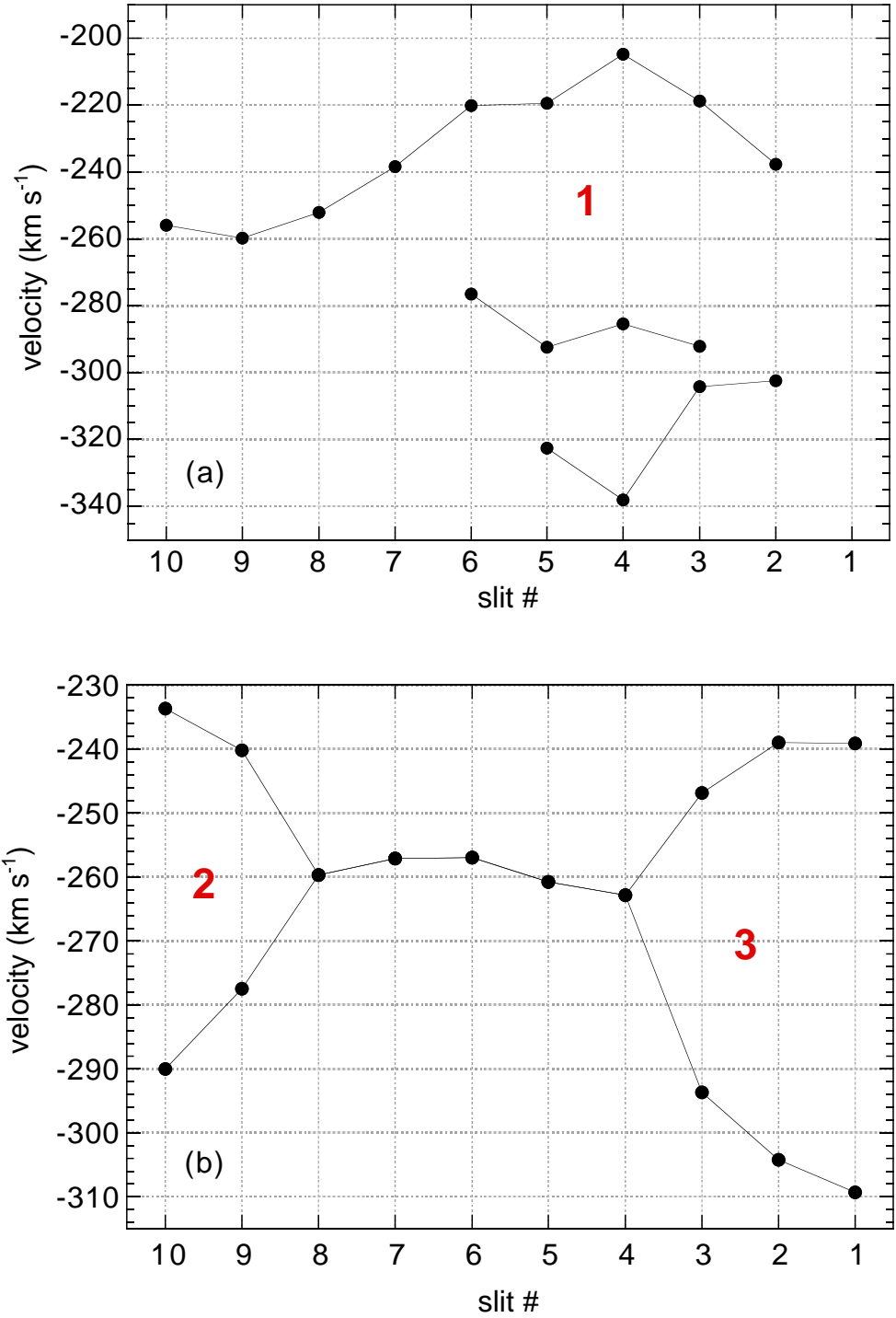}
\figcaption{The ongoing blowout. Kinematic signatures of the three main 
disrupted cavities discussed in the text.}
\end{figure}

\end{document}